\documentclass{jfm}
\usepackage{graphicx}
\usepackage{epstopdf, epsfig}
\usepackage{xcolor}
\usepackage{lineno}

\newcommand{\iu}{{\mathrm{i}\mkern1mu}}
\renewcommand{\Re}{{\mathit{Re}}}

\newcommand{\Pm}{{\mathit{Pm}}}

\newcommand{\Ha}{{\mathit{Ha}}}

\shorttitle{Mode selection for AMRI by thermal convection}
\shortauthor{A. Mishra, G. Mamatsashvili, M. Seilmayer, F. Stefani}

\title{One-winged butterflies: mode selection for azimuthal
magnetorotational instability by thermal convection}

\author{Ashish Mishra\aff{1,2}
  \corresp{\email{a.mishra@hzdr.de}},
  George Mamatsashvili\aff{1,3},
  Martin Seilmayer\aff{4}
 \and Frank Stefani\aff{1}}

\affiliation{\aff{1}Helmholtz-Zentrum Dresden-Rossendorf, Bautzner Landstr. 400, D-01328 Dresden, Germany
 \aff{2}Center for Astronomy and Astrophysics, ER 3-2,
 TU Berlin, Hardenbergstr. 36, 10623 Berlin, Germany
\aff{3}Abastumani Astrophysical Observatory, Abastumani 0301, Georgia
\aff{4}Staatliche Studienakademie Bautzen, L\"obauer Str. 1, 02625 Bautzen, Germany}

\begin{document}

\maketitle

\begin{abstract}
The effects of thermal convection on turbulence in accretion discs, and particularly its interplay with the magnetorotational instability (MRI), are of significant astrophysical interest. Despite extensive theoretical and numerical studies, such an interplay has not been explored experimentally. We conduct linear analysis of the azimuthal version of MRI (AMRI) in the presence of thermal convection and compare the results with our experimental data published before.  We show that the critical Hartmann number ($Ha$) for the onset of AMRI is reduced by convection. Importantly, convection breaks symmetry between $m = \pm 1$ instability modes ($m$ is the azimuthal wavenumber). This preference for one mode over the other makes the AMRI-wave appear as a ``one-winged butterfly''. 
\end{abstract}

%%%===============================================%%%
%%%===============================================%%%
\vspace{-4em}
\section{Introduction}\label{sec:introduction}

%%%===============================================%%%
%%%===============================================%%%

Magnetic processes are ubiquitous in astrophysics.
Magnetorotational instability \citep[MRI,][]{Balbus_Hawley_1991} is one of the most important candidates for explaining enhanced transport of angular momentum in accretion discs and mass concentrations onto the central object. MRI may also be non-linearly interwoven with the magnetic dynamo process, leading to the concept of {\it MRI dynamos} -- a class of instability-driven dynamos \citep{Rincon2019,Mamatsashvili_etal2020,Held_Mamatsashvili2022}. 

Since its rediscovery in 1991, there have been significant experimental efforts to study MRI in the laboratory. The PROMISE experiment, using GaInSn as a liquid metal, observed both the helical MRI \citep[with an imposed helical magnetic field,][]{Hollerbach_Rudiger2005, Stefani_etal2006} and the azimuthal MRI \citep[with an imposed azimuthal magnetic field,][]{Hollerbach_etal2010, Seilmayer_etal2014}, which both represent inductionless variants of MRI. By contrast, a conclusive confirmation of the standard MRI in the presence of an axial magnetic field is still elusive, despite promising recent findings \citep{Wang_etal2022b_PRL}. For this purpose, a large-scale sodium experiment is currently under construction in the frame of the DRESDYN project \citep{Stefani_etal2019}, aiming to reach large enough values of Lundquist and magnetic Reynolds numbers, $\sim 10$ and $\sim 40$, respectively, which are necessary for the onset and development of the standard MRI  \citep{Mishra_etal_SMRI_2022PhRvF, Mishra_etal_NonlinSMRI2023PhRvF}. 

While many numerical studies have shown that convection can foster hydrodynamic and magnetohydrodynamic turbulence in accretion discs, thereby enhancing accretion rate and angular momentum transport efficiency \citep{Klahr_Henning_Kley_1999ApJ, Bodo_etal2013, Coleman_etal2018ApJ,Held_Latter_2018MNRAS, Held_Latter_2021MNRAS},
there have been no endeavors to explore the interaction between MRI and convection in a laboratory setting. Here, in a first-of-its-kind attempt, we theoretically and experimentally study the azimuthal version of MRI (AMRI) in the presence of a {\it radial} temperature gradient which, although being different from the {\it vertical} stratification often considered in accretion discs, can still provide physical insights into the interplay between MRI and convection.

%%%===============================================%%%
%%%===============================================%%%

AMRI  is a non-axisymmetric instability arising in the presence of a purely azimuthal magnetic field and is characterized by dominant azimuthal wavenumbers $m=\pm 1$ \citep{Hollerbach_etal2010}. It emerges as a travelling wave in a differentially rotating flow that is otherwise hydrodynamically stable. AMRI was first observed in the PROMISE experiment \citep{Seilmayer_etal2014} as a characteristic travelling wave pattern, in a reasonable agreement with theoretical predictions. 
After improving the symmetry of the applied azimuthal field, the strongly interpenetrating waves still observed in the 2014 experiment
were replaced by a much clearer ``butterfly''-like wave pattern, revealing, however, a new noteworthy effect of symmetry breaking between the two unstable $m=\pm 1$ modes \citep{Seilmayer_2016}, the reason for which was not clear by that time. In a more recent linear study of AMRI \citep[hereafter Paper I]{Mishra_etal_AMRI_2021JFM},  we showed that the absolute form of AMRI with zero group velocity (but nonzero phase velocity), which is more relevant and important in experiments, successfully describes the observed butterfly-shaped structure of axially upward and downward traveling waves. 

The motivation for the present  study comes from the recent work by \cite{Seilmayer_etal2020}, who experimentally investigated  the interplay of AMRI and thermal convection in PROMISE. They observed that convection driven by radiative heat flux from the central current-carrying rod causes the symmetry breaking of the $m=\pm 1$ AMRI waves and the systematic shift of their characteristic frequencies, phase velocities and wavenumbers. Moreover, the direction of the phase velocity of the dominant AMRI wave appeared to be linked to the direction of heat flux defining the convective motion. Our goal is to explain this behavior based on the linear stability analysis of a dissipative Taylor-Couette (TC) flow in the presence of thermal convection and an azimuthal background magnetic field. Following Paper I, in this work we also focus on the absolute form of AMRI. The main result is that the convection flow causes symmetry breaking between the $m=\pm 1$ AMRI modes, giving preference to either of these two modes and increasing its growth rate, while decreasing that of another. This preferred mode gives rise to a characteristic ``one-winged butterfly'' pattern of the AMRI wave observed in the experiments.
%\vspace{-2em}
%%%==============================================%%%
%%%==============================================%%%

\section{Theoretical model} \label{sec:mathematicalformulation}

%%%===============================================%%%
%%%===============================================%%%

\begin{figure}
    \centering
    \includegraphics[width=0.35\textwidth]{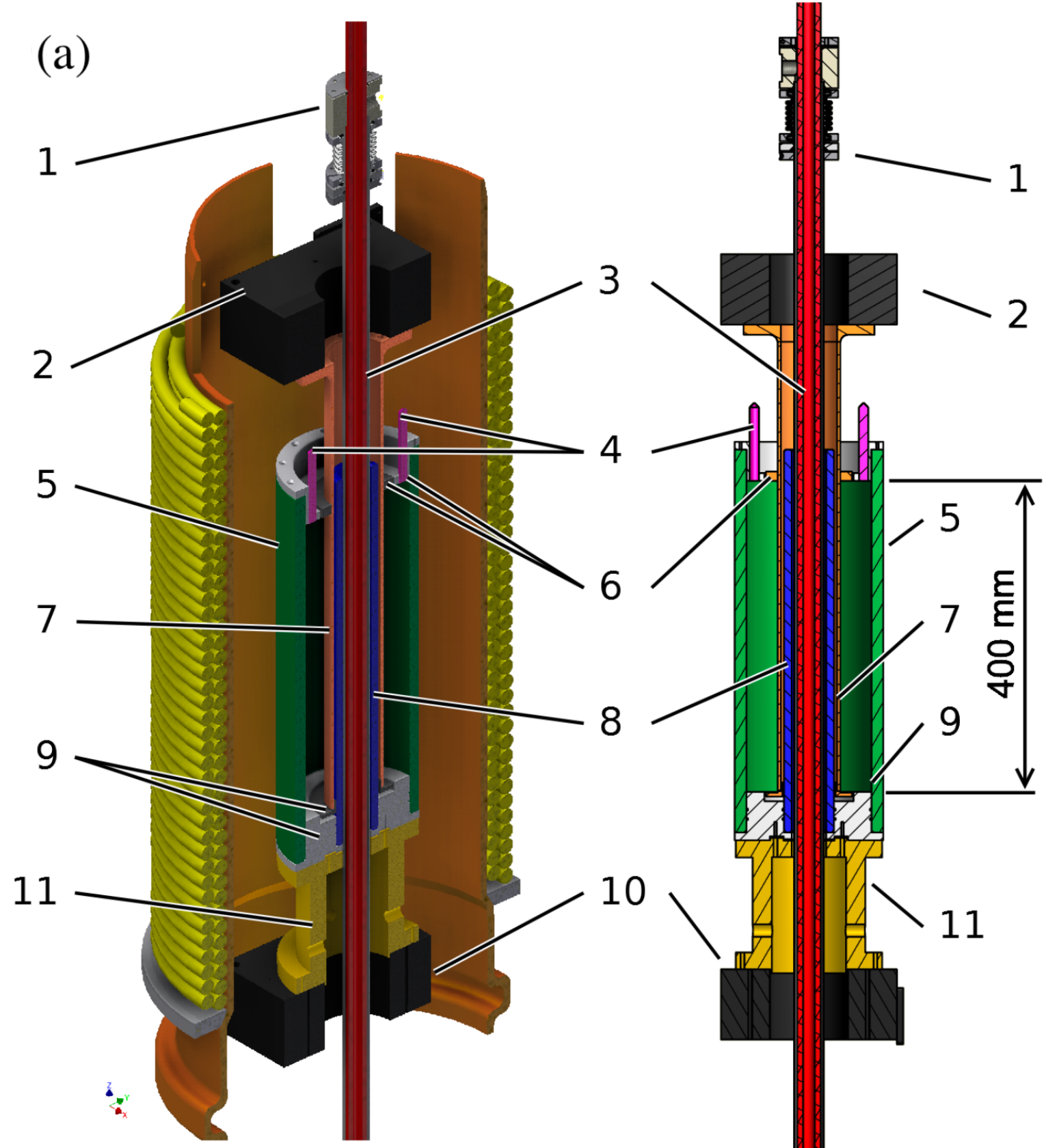}
    \hspace{5em}
    \includegraphics[width=0.35\textwidth]{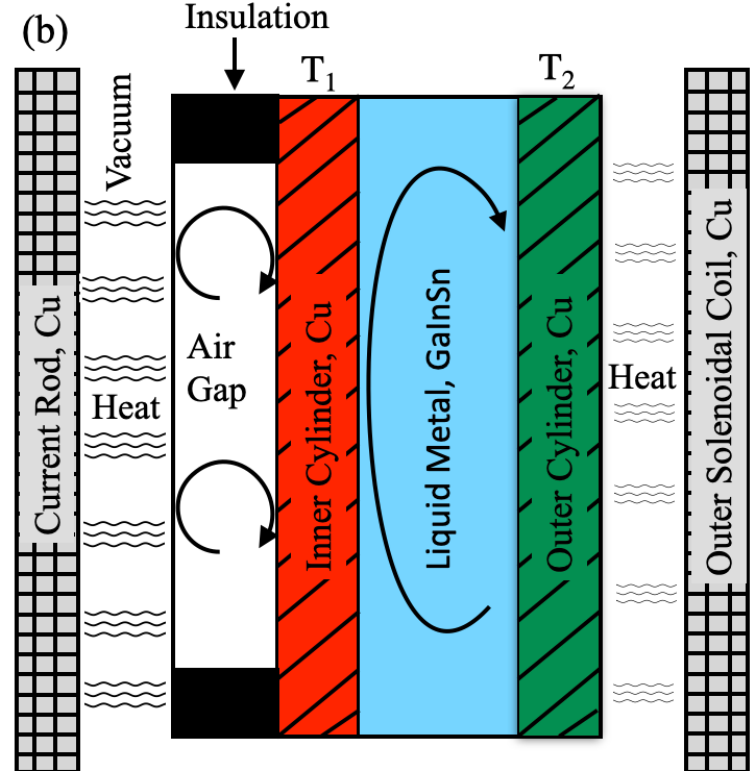}
    \caption{The PROMISE experiment using GaInSn as a working fluid. (a) Cross-section of the experiment with dimensions \({h=40}\,\mathrm{cm}\), \({r_\mathrm{in} = 4}\,\mathrm{cm}\) and \({r_\mathrm{out} = 8}\,\mathrm{cm}\): (1) vacuum insulation, (2) upper motor, (3) current-carrying copper rod, (4) UDV sensors, (5) outer cylinder, (6) top acrylic glass split rings, (7) inner cylinder, (8) central cylinder, (9) bottom split rings, (10) bottom motor and (11) interface. 
    (b) 2D sketch of the PROMISE-TC setup showing heat radiation from the vacuum-insulated current-carrying rod. Heat flux is directed from the inner to outer cylinder with temperatures $T_1$ and $T_2$, respectively, obeying $T_1 > T_2$, which induces convective motion in the fluid.  A reverse temperature gradient and hence opposite convective velocities can be set by pre-heating the outer solenoidal coil before starting the experiment.}
    \label{fig:tcsetup}
\end{figure}

We consider an infinitely long cylindrical TC flow setup consisting of inner and outer cylinders with radii $r_\mathrm{in}$ and $r_\mathrm{out}$ rotating with angular velocities $\Omega_\mathrm{in}$ and $\Omega_\mathrm{out}$ in the cylindrical coordinates $(r, \phi, z)$ corresponding to the PROMISE setup (figure \ref{fig:tcsetup}). 
In the absence of endcaps, the equilibrium azimuthal flow $u_{0\phi}=r\Omega(r)$ between the cylinders has a classical hydrodynamical TC profile of angular velocity $\Omega(r)=S_1+S_2/r^2$, where
$S_1=\Omega_\mathrm{in}(\mu-\eta_{\Omega}^2)/(1-\eta_{\Omega}^2)$ 
and $S_2=\Omega_\mathrm{in}(1-\mu)r_\mathrm{in}^2/(1-\eta_{\Omega}^2)$ with the ratio of the cylinders' radii $\eta_{\Omega}=r_\mathrm{in}/r_\mathrm{out}$ ($=0.5$ in PROMISE) and angular velocities $\mu=\Omega_\mathrm{out} / \Omega_\mathrm{in}$. Note that a split endcaps configuration is used in the PROMISE setup, which significantly reduces the global Ekman pumping, thereby sustaining this TC profile in the bulk flow to a good approximation \citep{Stefani_etal2009}. 
A central rod carrying current $I$ produces an azimuthal magnetic field $B_{0\phi}(r)=\mu_0I/(2\pi r)$ between the cylinders, where $\mu_0$ is the magnetic permeability of vacuum. Since this current of the order of $10~\rm kA$ produces appreciable Ohmic dissipation, $P_\mathrm{rod}\sim 1~\rm kW$, a water cooling system and a vacuum insulation balance the thermal heating at the centre, see figure \ref{fig:tcsetup}.

The MHD equations for an incompressible fluid with a temperature gradient are
\begin{equation} \label{equation_momentum}
	\frac{\partial {\boldsymbol u}}{\partial t}+({\boldsymbol  u}\cdot\nabla){\boldsymbol u}=-\frac{1}{\rho} \nabla p + \frac{{\boldsymbol J}\times {\boldsymbol B}}{\rho}+\nu \nabla^2 {\boldsymbol u} -{\boldsymbol g}\beta \delta T ,
\end{equation}

\begin{equation}	\label{equation_induction}
	\frac{\partial \boldsymbol{B}}{\partial t}=\nabla \times(\boldsymbol{u}\times \boldsymbol{B})+\eta \nabla^2 \boldsymbol{B},	
\end{equation}

\begin{equation}\label{equation_div_free}
	\nabla \cdot \boldsymbol{u}=0, ~~~~ \nabla \cdot \boldsymbol{B}=0 .
\end{equation}

\noindent
where $\rho$ is the constant density, $\boldsymbol{u}$ is the velocity, $p$ is the thermal pressure, $\boldsymbol{B}$ is the magnetic field, $\nu$ and $\eta$ are, respectively, the fluid viscosity and magnetic diffusivity and $\boldsymbol{ J}=\mu_0^{-1}\nabla\times\boldsymbol{B}$ is the current density. The last term  $-\boldsymbol{g}\beta \delta T$ on the right hand side of equation (\ref{equation_momentum}) is the buoyancy force in the Boussinesq  approximation driving thermal convection flow \citep{Landau}, where $\beta>0$ is the coefficient of thermal expansion of the fluid, $\delta T=T-T_0$ is the temperature deviation from the reference stationary profile $T_0(r)$ in the absence of convection, which is set by the temperatures $T_1$ and $T_2$ of the inner and outer cylinders, respectively (figure \ref{fig:tcsetup}). The gravitational acceleration $\boldsymbol{g}=-g\boldsymbol{e}_z$ points opposite the unit vector $\boldsymbol{e}_z$ of the $z$-axis. The background azimuthal magnetic field due to the central current $I$ is written as $\boldsymbol{B_0} = B_0(r_\mathrm{in}/r)\boldsymbol{e}_\phi$, where $B_0= \mu_0I/(2\pi r_\mathrm{in})$ is the value of this field at $r_\mathrm{in}$ and $\boldsymbol{e}_\phi$ is the unit vector in the azimuthal direction. In addition, the present setup of the PROMISE experiment uses an enhanced pentagonal-shaped frame system maintaining an axisymmetry of the background azimuthal field with a relative error $\Delta B_{\phi} / B_{0} < 10^{-2}$. 

Apart from being a magnetic field source, the current also represents a heat source, as illustrated in figure \ref{fig:tcsetup}(b), where the thermal radiation from the Joule heating of current-carrying central rod transports heat outwards to the inner cylinder \citep[see details in][]{Seilmayer_etal2020}. As the inner cylinder's wall heats up, it drives convective motion of the fluid with an axial velocity $u_{0z}$, which is directed upwards along that wall, but downwards along the outer cylinder wall.  A heat flux in the opposite direction and hence a reverse convection flow is obtained by pre-heating the outer solenoidal coil (e.g., by letting current through it) before the start of the experiment. Assuming balance between Lorentz and axial buoyancy forces in a stationary convection flow, \citet{Seilmayer_etal2020} estimated the characteristic  axial velocity of this flow as $u_{0z}\approx 0.2 ~ \rm mm \cdot s^{-1}$ at the outer cylinder for a temperature difference $\Delta T\sim 0.1~\rm K$ and a current of $I=20~\rm kA$ used in those experiments. This velocity is smaller than that of the basic aizmuthal TC flow. Thus, the equilibrium flow represents a combination of the main TC flow and the radially varying axial velocity $u_{0z}$, i.e., $\boldsymbol{u}_0=(0, r\Omega(r), u_{0z}(r))$. Due to small temperature difference, we neglect the thermal effects (i.e., buoyancy term) for \textit{perturbations} analyzed in the next section.  In fact, as we will see below, thermal convection influences the dynamics of AMRI primarily through its axial velocity.
%%%===============================================%%%
%%%===============================================%%%

\subsection{1D linear stability analysis}

%%%===============================================%%%
%%%===============================================%%%
We consider small perturbations of velocity ${\boldsymbol u}'={\boldsymbol u}-{\boldsymbol u}_0$, pressure $p'=p-p_0$, and magnetic field ${\boldsymbol b}'={\boldsymbol B}-{\boldsymbol B}_0$ about the above equilibrium flow, which are functions of the radius $r$ and depend on $t, \phi$ and  $z$ as a normal mode $\propto {\rm exp}(\gamma t+\iu m\phi+\iu k_z z)$, where $\gamma$ is the (complex) eigenvalue, while $k_z$ and the integer $m$ are the axial and azimuthal wavenumbers, respectively.  A positive real part (growth rate) of any eigenvalue,  ${\cal R}(\gamma) > 0$,  indicates the istability of perturbations.
We normalise length by $r_\mathrm{in}$, time by $\Omega_\mathrm{in}^{-1}$, $\gamma$ and $\Omega(r)$ by $\Omega_\mathrm{in}$, ${\boldsymbol u}$ by $\Omega_\mathrm{in} r_\mathrm{in}$, $p$ by $\rho r_\mathrm{in}^2\Omega_\mathrm{in}^2$, ${ \boldsymbol B_0}$ by $B_0$, and ${ \boldsymbol b}'$ by $\Re\Pm\cdot B_0$, where $\Re = \Omega_\mathrm{in}r_\mathrm{in}^2/\nu$ is the Reynolds number and the magnetic
Prandtl number $Pm = \nu/\eta=1.4 \times 10^{-6}$ is very small typical of the working liquid GaInSn in the experiments. We also define another main parameter -- the Hartmann number $\Ha = B_0 r_\mathrm{in}/\sqrt{\rho\mu_0\nu\eta}$ ($\approx 7.77 \cdot I $\,/kA for GaInSn) characterizing magnetic field strength.   

The perturbations of velocity and magnetic field are divergence-free (equation \ref{equation_div_free}),  so that they can be split  into toroidal and poloidal components (primes will be omitted), $\boldsymbol{u}=\nabla\times (e{\bf e}_r)+\nabla\times\nabla\times (f{\bf e}_r)$, $\boldsymbol{b}=\nabla\times (g{\bf e}_r)+\nabla\times\nabla\times (h{\bf e}_r)$, where $e,f,g,h$ are the functions of only radius, $\boldsymbol{e}_z$ is the radial unit vector and $\nabla=(\partial/\partial r, im/r, ik_z)$. Linearizing equations (\ref{equation_momentum})–(\ref{equation_induction}), substituting these representations of the velocity and magnetic field, ignoring the buoyancy force and using the above normalizations, we finally get a system of four coupled 1D linear eigenvalue equations \citep{Hollerbach_etal2010},
\begin{eqnarray}\label{equations_eigenvalue}
	\Re\gamma(C_2 e+C_3 f)+C_4 e+C_5 f& = &\Re (E_1+F_1) + \Ha^2 (G_1 + H_1),\\
	\Re\gamma(C_3 e+C_4 f)+C_5 e+C_6 f& = &\Re (E_2 + F_2) + \Ha^2 (G_2 + H_2),\\
	\Re\Pm\gamma(C_1 g+C_2 h)+C_3 g + C_4 h& = &E_3+F_3 + \Re\Pm (G_3 + H_3),\\
	\Re\Pm\gamma(C_2 g + C_3 h)+C_4 g+C_5 h& = &E_4+F_4 + \Re\Pm (G_4 + H_4),
\end{eqnarray}
\noindent
 where the operators $C_n$ are given by $C_n \cdot={\bf e}_r(\nabla\times)^n(\cdot~{\bf e}_r), ~n\in [1,6]$, while other terms on the right hand side of these equations are
 \[
 	E_1=-\iu m \mathrm{\Delta} \Omega e-\textcolor{blue}{\iu k_z u_{0z} \mathrm{\Delta} e}, ~~
 	E_2=\iu k_z  \mathrm{\Delta} \Omega e + \textcolor{blue}{2\iu mk_z^2 u_{0z}e/r^2},~~
 	E_3=0, ~~
 	E_4=imr^{-2} \Delta e, 
 \]
 \[
 	F_1=\iu k_z (\mathrm{\Delta} \Omega+\mathrm{\Delta} r\Omega')f+ \textcolor{blue}{\iu m (2k_z^2 u_{0z}/ r-\mathrm{\Delta} u_{0z}')f/r},
 \]
 \[
 	F_2= -\iu m [\Omega(C_4+4k_z^2/r^2)+\Delta((r^2\Omega')'/r^2)]f  -\textcolor{blue}{\iu k_z[u_{0z}C_4+\Delta(r(u_{0z}'/r)')]f},\\
 \]
 \[
 	F_3=im r^{-2} \Delta f,\\ ~~~~
 	F_4=-ik_zr^{-2} \hat{\Delta} f ,\\
\]
\[
 	G_1=imr^{-2} \Delta g, ~~~~
 	G_2=-ik_z r^{-2} \hat{\Delta}g, ~~~~
 	G_3=0, ~~~~
 	G_4=-\iu m \mathrm{\Delta} \Omega g-\textcolor{blue}{\iu k_z \mathrm{\Delta} u_{0z} g}, 
 \]
 \[
 	H_1=-2im^2 k_z r^{-4} h, ~~~
 	H_2=im r^{-2} C_4 h+4im k_z^2 r^{-4} h,~~~
 	H_3=-\iu m \mathrm{\Delta} \Omega h-\textcolor{blue}{\iu k_z u_{0z} \Delta h}, 
 \]
 	\[
 	H_4=\iu k_z (2m^2 r^{-2} \Omega-\mathrm{\Delta} r\Omega' )h+\textcolor{blue}{\iu m (2 k_z^2 u_{0z}/ r + \mathrm{\Delta} u_{0z}' )h/r},\\
 \]
where $\hat{\Delta}=4m^2 r^{-2}+2k_z^2$ and $\Delta = m^2 r^{-2}+k_z^2$. The stationary and radially varying axial velocity $u_{0z}(r)$ induced by convection introduces new contributions in several terms $E_1, E_2, F_1, F_2, G_4, H_3, H_4$ compared to the case without convection $(u_{0z}=0)$, which are highlighted in blue. The inner and outer cylinders of the PROMISE device are made of copper (figure \ref{fig:tcsetup}) and hence conducting boundary conditions are used for the magnetic field, while no-slip conditions for the velocity. The eigenvalue problem posed by equations (\ref{equations_eigenvalue})-(2.7) together with these boundary conditions allow us to determine the eigenvalues $\gamma$ and the associated eigenmodes. To solve this problem, we use the 1D code of \citet{Hollerbach_etal2010} based on the spectral collocation method with $N = 30–40$ Chebyshev polynomials, thereby reducing these linear differential equations to a large $4N \times 4N$ matrix eigenvalue equation for $\gamma$ which is then solved with LAPACK library.

As already noted in Introduction, following Paper I, in this work we focus on the absolute form of AMRI and, using the procedure described in that paper, identify the corresponding unstable modes that are characterized by zero group velocity (but nonzero phase velocity) in the axial $z$-direction and hence stay inside the TC device. A key feature of these modes is that their axial wavenumber is complex $k_{z,a}={\cal R}(k_{z,a})+i{\cal I}(k_{z,a})$, resulting in the increase of mode amplitudes along or opposite the $z$-axis depending on the sign of the imaginary part of the wavenumber ${\cal I}(k_{z,a})$ (see details in Paper I). The requirement of zero group velocity implies mathematically that the wavenumber $k_{z,a}$ should be a saddle point of the dispersion relation $\gamma(k_z)$, where its complex derivative is zero, $\partial \gamma (k_z)/\partial k_z|_{k_z=k_{z,a}}=0$. From this condition we determine $k_{z,a}$ and then calculate the corresponding growth rate $\gamma_a={\cal R}(\gamma (k_{z,a}))$ for different $Ha$ and $Re$.

Note that the axial velocity of convection, $u_{0z}$, its radial shear, $u_{0z}'$, and the azimuthal wavenumber $m$ enter equations (\ref{equations_eigenvalue})-(2.7) as products $mu_{0z}$ and $mu_{0z}'$ in the right hand side terms of these equations, which thus introduce asymmetry with respect to $m$ -- the selection of $m=1$ or $m=-1$ AMRI-unstable modes. As is shown below, this selection effect primarily manifests itself in the enhancement of the growth rate of either of these two modes and lowering (or even suppressing) another.

\subsection{Temperature profile}

\begin{figure}
	\centering
	\includegraphics[width=0.3\textwidth]{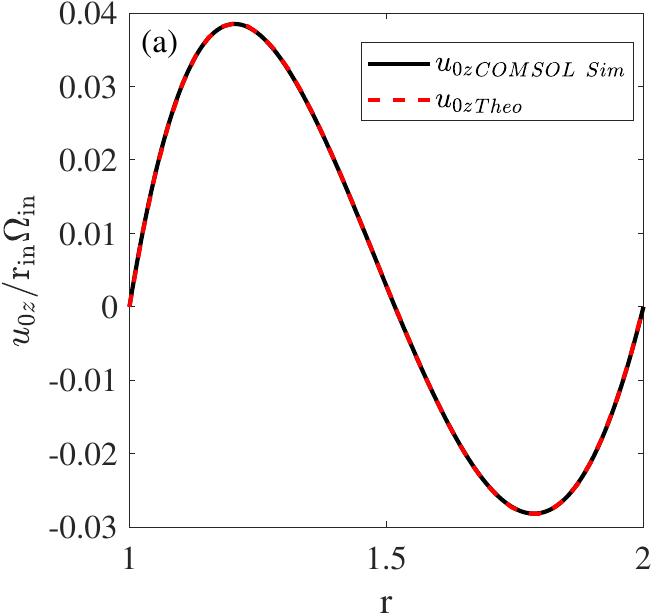}
	\hspace{3em}
	\includegraphics[width=0.29\textwidth]{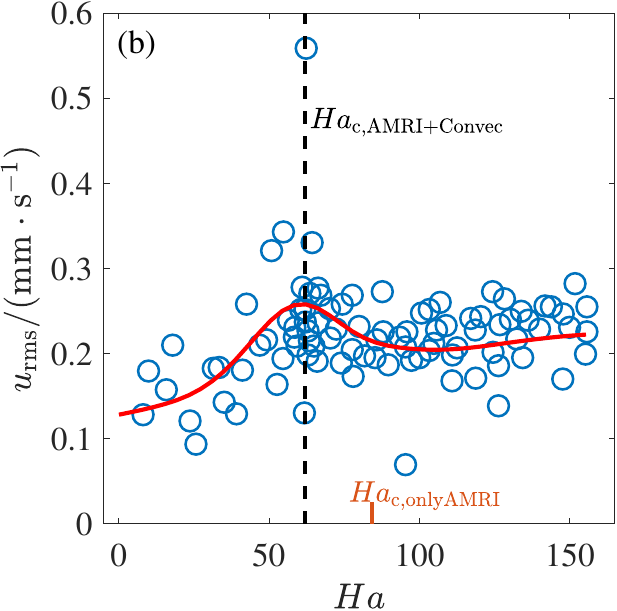}
	\caption{(a) Axial velocity $u_{0z}$ of convection vs.  $r$ obtained when heat flux is directed from the inner to outer cylinder from the axisymmetric ($m=0$) COMSOL simulations (black) for current $I=20~ \rm kA$ and from equation (\ref{convec_uz_velo}) (red) with the amplitude factor $A_{0z}\approx 4.25$. (b) The rms of axial velocity, $u_{z,\mathrm{rms}}$ (blue circles) calculated from its time- and azimuthally-averaged radial profile measured in the experiment as a function of $Ha$ (i.e.,  current $I$). The red curve denotes a Gaussian fit applied to these data points. The vertical dashed line marks the critical $Ha_c\approx 62$ for the onset of AMRI with convection,  see also figure \ref{fig:amriconvectiongrowthvelo}(b).} \label{fig:amriconvectionu0zA0z}
\end{figure}

So far, the mathematical expression of the axial velocity $u_{0z}$ has been kept arbitrary,  i.e.,  in principle it may have any radial dependence and can, therefore, be explored over a broad range of parameter space in $B_0$ and $\Omega_\mathrm{in}$ which may be of astrophysical interest. However, finding an exact form of an established stationary $u_{0z}$ in the given cylindrical annulus of a TC flow bounded by endcaps due to the combined action of buoyancy, inertia, and the imposed azimuthal field  is another problem, which we do not address here. Instead, we adopt its simplified model -- infinitely long rotating cylinders in the presence of a radial  temperature gradient and a current-free azimuthal field.  Since $Pm \ll 1$, the effect of Lorentz force on the convective motion is small. In this case, an exact stationary axisymmetric solution of momentum equation (\ref{equation_momentum}) was found by \citet{Ali_Weidmann1990JFM} that consists of a TC flow $u_{0\phi}=r\Omega(r)$ and the convection axial velocity given by
\begin{equation}\label{convec_uz_velo}
	u_{0z}=\frac{A_{0z}}{(1-\eta_\Omega)^2}\Biggl\{\Bigg(\frac{A}{B}\Bigg) \Bigg((1-\eta_\Omega)^2 r^2-1+(1-\eta_\Omega)^2C -\frac{1}{4}[(1-\eta_\Omega)^2r^2-\eta_\Omega^2]C\Bigg)\Biggr\},
\end{equation}
\noindent
where 
\[
A=(1-\eta_\Omega^2)[1-3\eta_\Omega^2-4\eta_\Omega^4 \ln (\eta_\Omega)], ~~~ B=16\bigl[(1-\eta_\Omega^2)^2+(1-\eta_\Omega^4) \ln(\eta_\Omega)\bigr], ~~~C=\frac{\ln(r/r_{out})}{\ln(\eta_\Omega)}
\]
and the amplitude factor $A_{0z}$ is a constant proportional to the temperature difference between the cylinders, $\Delta T=T_1-T_2$, which can be obtained either experimentally or via simulations of the corresponding real system of a bounded TC flow with radial temperature gradient and the magnetic field. To test the validity of this expression for the present setup, in figure \ref{fig:amriconvectionu0zA0z}(a) we compare the radial profiles of $u_{0z}$ as given by equation (\ref{convec_uz_velo}) and that obtained from the simulations of the real system at the central current $I=20 ~ \rm{kA}$ using COMSOL Multiphysics software. For this value of the current, the analytical solution (\ref{convec_uz_velo}) reproduces well the radial profile of the axial velocity from the simulations if we choose $A_{0z} \approx 4.25$. 

We prefer,  however,  to directly derive $A_{0z}$ from the experimental data, as these simulations are somewhat limited in several respects (low resolution, only axisymmetric, etc.) and, in particular, do not include AMRI.   Figure \ref{fig:amriconvectionu0zA0z}(b) shows the rms calculated from the time- and azimuthally-averaged axial velocity $u_{z,\mathrm{rms}}$ measured in the experiment close to the outer cylinder where sensors are located.  We apply a Gaussian fit to these data and, using those fitted values in equation (\ref{convec_uz_velo}) at the radius of the sensor locations, determine the corresponding $A_{0z}$ (which in fact differs from $u_{z,\mathrm{rms}}$ at that radius by a constant factor) for a given $Ha$.  From the dependence of $u_{z,\mathrm{rms}}$ on $Ha$ in figure \ref{fig:amriconvectionu0zA0z}(b) it is seen that the magnetic field perturbation growing as a result of AMRI back-reacts on (slows down) the convection due to nonlinearity, modifying the dependence of the resulting rms of axial velocity on $Ha$ (denoted by the black dashed line), which would otherwise increase nearly proportional to $Ha$ \citep{Seilmayer_etal2020}. Afterwards, using those values of $A_{0z}$ obtained from the experimental data back into equation (\ref{convec_uz_velo}), we recover the entire radial profile of $u_{0z}(r)$ and plug it into the various terms on the right hand side of the eigenvalue equations (\ref{equations_eigenvalue})-(2.7). Ideally, one would treat the dynamics of AMRI and the background axisymmetric convection flow self-consistently, that is, taking into account the mutual nonlinear interaction of the AMRI wave and the basic convection flow via Lorentz force, which will be the subject of future more extensive analysis. Here,  we focus instead only on the \textit{linear} dynamics of AMRI upon the established convection flow, whose axial velocity amplitude $A_{0z}$ has been directly derived from the experiments.

%\vspace{-1em}
%%%===============================================%%%
%%%===============================================%%%
\section{Comparison of experimental and theoretical results}
%%%===============================================%%%
%%%===============================================%%%

\begin{figure}
	\centering
	\includegraphics[width=0.7\textwidth]{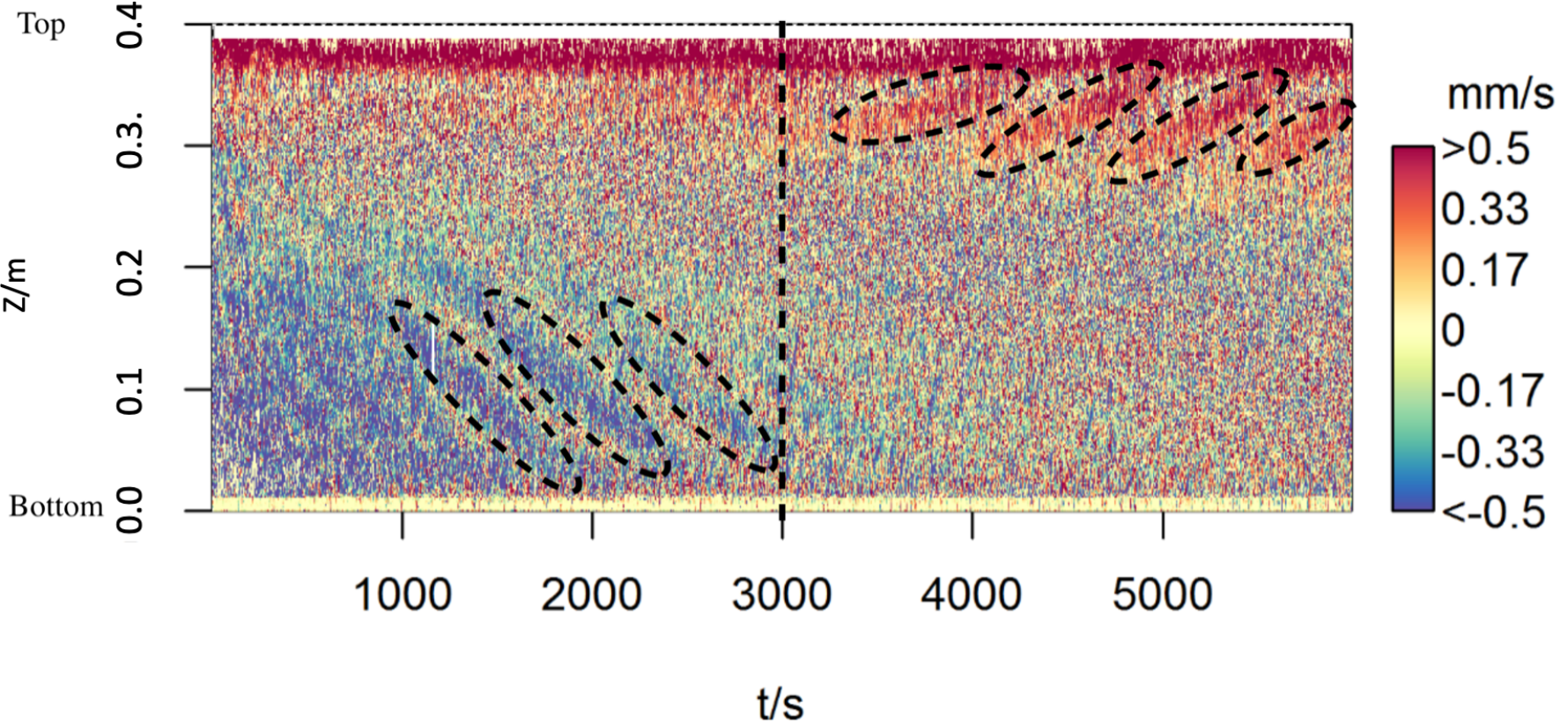}
	\caption{AMRI wave in the presence of thermal convection. UDV raw data of the axial velocity $u_{z}$ measured by the sensor close to the outer cylinder as 2D series in $t$ and $z$ at current $I = 12.87~ \rm kA$ ($Ha = 100$) and $Re=1480$. The dominant direction of the AMRI wave (marked by dashed elliptical curves) depends on the direction of heat flux, which is initially from the outer to inner cylinder up to $t = 3000~s$ (marked by black dashed line) and then,  when the outer coil has cooled down, heating from the central rod prevails, switching the direction of the heat flux.  This figure is adopted from \citet{Seilmayer_etal2020}.}\label{fig:AMRI_2020paper}
\end{figure}

To experimentally study the effect of convection on AMRI, an upgraded PROMISE setup (figure \ref{fig:tcsetup}) was used at fixed $\Omega_\mathrm{in}=2\pi\times 0.05~\rm Hz$, yielding  $Re=1480$, and $\mu=0.26$, but different $Ha$, as indicated in figure \ref{fig:amriconvectionu0zA0z}(b) \citep[see also][]{Seilmayer_etal2020}. We note that the PROMISE facility was not originally designed to conduct experiments with thermal convection. This  has introduced some limitations to the experiments, for example, as the source of heat is the current-carrying rod rather than a special heating device, the typical temperature gradient between the cylinders, as noted above, is small ($\Delta T \sim 0.1 \rm K$) and not adjustable, leading accordingly to small axial velocities. Nevertheless,  given the working liquid GaInSn, it turned out that even such a small temperature gradient leads to a sufficient convective flow velocity that can be observed and have an effect on AMRI \citep{Seilmayer_etal2020}. Indeed, it was found in those experiments that, contrary to the case of AMRI without convection, where the upward and downward moving waves appear \textit{symmetrically} in the whole cylinder height (see figure 5 in Paper I), thermal convection breaks this symmetry between the $m=\pm 1$ modes, resulting in the AMRI waves to appear either in the upper or lower half of the  cylinder depending on the direction of the heat flux, as it is seen from the variation of the axial velocity $u_z$ in the $(t,z)$-plane, also known as Hovm\"oller, or ``butterfly'' diagram, presented in figure \ref{fig:AMRI_2020paper}, which has been adopted from the experimental work of \cite{Seilmayer_etal2020}.  As we will show below, the effect of the heat flux direction on the wave pattern structure (butterfly diagram) is, however, indirect -- it is the induced convection flow velocity $u_{0z}$ and its radial shear that primarily causes symmetry breaking between these two modes. Moreover, as we will see, convective flow can also cause a shift in phase velocities and onset threshold of AMRI.

We can view the selection of AMRI modes also in analogy with the solar dynamo. In figure \ref{fig:AMRI_2020paper}, the heat flux is initially from the outer to inner cylinder, i.e., fluid is rising ($u_{0z}>0$) near the outer and sinking ($u_{0z}<0$) near the inner cylinder. Since the axial velocity at the inner cylinder is (about 1.33 times) higher than that at the outer one (see figure \ref{fig:amriconvectionu0zA0z}a), the former prevails in carrying the AMRI wave -- the direction of the wave phase velocity coincides with that of the downward convection velocity near the inner wall. This effect is quite similar to that of the stronger equator-ward meridional circulation in the solar tachocline which governs the direction of the butterfly diagram of sunspots. When the heat flux is from the inner to the outer cylinder at later times in figure \ref{fig:AMRI_2020paper}, the fluid rises near the inner cylinder and carries the AMRI wave upwards.

\begin{figure}
	\centering
	\includegraphics[width=0.6\textwidth]{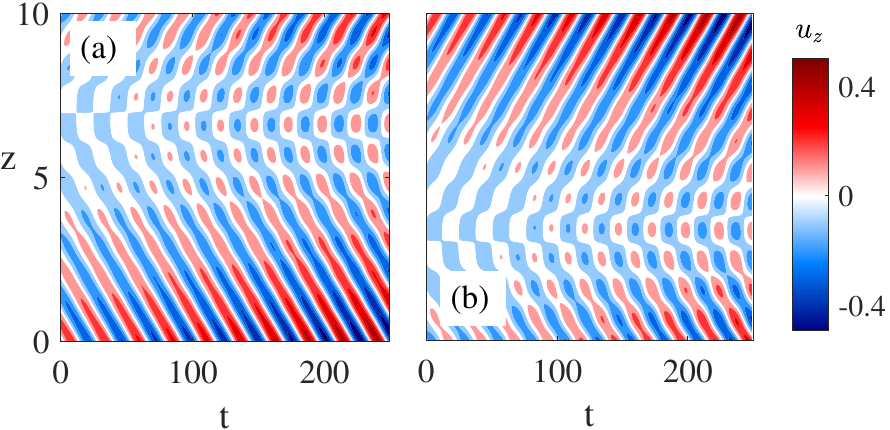}
	\caption{Spatio-temporal variation (Hovm\"oller, or butterfly diagram) of the perturbed axial velocity $u_z$ of the most unstable AMRI wave in the $(t, z)$-plane at $I=13 ~ \rm kA$ ($\Ha \approx 101$) and $Re=1480$. (a) $m=-1$ AMRI mode dominates at the bottom when the heat flux is directed from the outer to inner cylinder,  while (b) $m=1$ AMRI mode dominates at the top when the heat flux is directed from the inner to outer cylinder. This clearly shows symmetry breaking, or selection effect between the $m=\pm 1$ modes due to convection. This selection between these two modes depends on the direction of convective flow linked to the heat flux, which is consistent with the experimental findings in figure \ref{fig:AMRI_2020paper}.} \label{fig:amriconvectionvelocity}
\end{figure}

To interpret the behaviour seen in figure \ref{fig:AMRI_2020paper} on a more physical footing, we conduct the 1D linear stability analysis for the experimental parameters given above. In this case, the AMRI modes have $m=\pm 1$ while other MHD or hydrodynamic higher $|m|\geq 2$ modes are stable. 
Figure \ref{fig:amriconvectionvelocity} shows a similar butterfly diagram of the perturbed axial velocity $u_z$ associated with the AMRI wave in the $(t,z)$-plane in the presence of convection at $I = 13~\rm kA$ ($Ha=101$) 
for both, from the outer to inner cylinder and vice versa, directions of the heat flux. For only AMRI without convection considered in Paper I, the spatio-temporal variation of axial velocity exhibits a pattern of upward and downward moving waves in the $(t,z)$-plane, which contains \textit{both} $m=\pm 1$ modes with equal weights (growth rates) located symmetrically with respect to the mid-plane of the cylinder (see figure 5 in Paper I).  As shown in that paper, this is due to opposite signs (but equal absolute values) of the imaginary parts of the complex axial wavenumbers -- ${\cal I}(k_{z,a})<0$ for $m=1$ and ${\cal I}(k_{z,a})>0$ for $m=-1$ -- of these two absolute AMRI modes, which hence appear to be concentrated towards the top and bottom ends of the cylinders, respectively, but both components are always present at the same time due to axial symmetry. The direction of the phase velocities coincides with the corresponding direction of the wave concentration.

The dynamics is significantly altered in the presence of radial temperature gradient. The symmetry between the $m = \pm 1$ modes is broken due to  convection velocity $u_{0z}$ -- either of these two modes is preferred over another depending on the direction of $u_{0z}$, as it is clearly seen in figure \ref{fig:amriconvectionvelocity}.  When the heat flux is directed from the outer to inner cylinder (i.e., rising $u_{0z}>0$ near the outer cylinder and sinking $u_{0z}<0$ near the inner one), $m =-1$ mode with ${\cal I}(k_{z,a})>0$ is dominant that increases opposite the $z$-axis and has its phase  velocity also directed opposite this axis (figure \ref{fig:amriconvectionvelocity}a). By contrast, for the reversed heat flux direction, $m = 1$ mode with ${\cal I}(k_{z,a})<0$ is dominant that increases along the $z$-axis and has its phase velocity also directed along this axis (figure \ref{fig:amriconvectionvelocity}b). As a result, for each direction of the heat flux, the butterfly diagram takes on a predominantly one-winged structure corresponding to the dominant mode, being mostly concentrated near the top (``upper wing'' for $m=1$) or bottom (``lower wing'' for $m=-1$) of the cylinder. This spatio-temporal variation of the wave velocity is in a qualitative agreement with the experimentally observed pattern of AMRI waves in figure \ref{fig:AMRI_2020paper}.

\begin{figure}
	\centering
	\begin{minipage}{0.90\textwidth}
		\includegraphics[width=\textwidth]{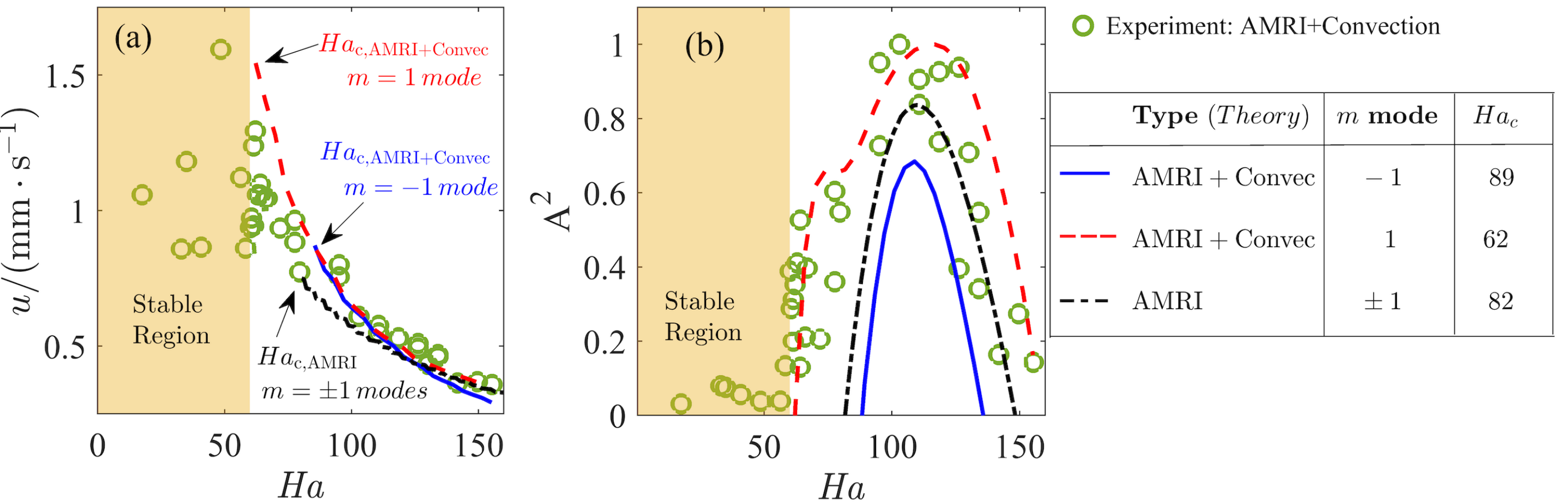}
	\end{minipage}
   \caption{ (a) Phase velocities and (b) normalized energy content $\rm A^2$ (in a.u.) that is assumed to be proportional to the growth rate $\gamma_a$ of the absolute AMRI in the presence of convection at $Re=1480$. The heat flux is directed from the inner to outer cylinder. Green circles denote the experimental data, red-dashed lines correspond to the dominant $m=1$ mode, while blue line to the subdominant $m=-1$ mode, implying symmetry breaking between these two modes in contrast to the case without convection (black dot-dashed line) where the growth rates of both $m=\pm 1$ AMRI modes are equal.  Theoretical values of $\rm A^2$ are normalized by its maximum for the dominant $m=1$ AMRI mode occurring at $Ha=116$ (corresponding to the highest growth rate $\gamma_{a,max}=0.0043$).  Shaded yellow regions are AMRI-stable.} \label{fig:amriconvectiongrowthvelo}
\end{figure}

Figure \ref{fig:amriconvectiongrowthvelo}(a) compares the phase velocities $u$ of the AMRI waves as a function of $Ha$ from the experiments and the linear analysis in the presence of heat flux from the inner to outer cylinder. In the experiment, $u$ is measured near the outer cylinder using the Ultrasound Doppler Velocimetry (UDV). It is seen that the theoretical values of the phase velocities for $m=\pm 1$ AMRI wave modes from the linear analysis match quite well the experimental ones. This demonstrates the deviation of the phase velocity from the pure AMRI wave without convection. Due to symmetry breaking, the $m=1$ mode has somewhat larger phase velocity than the $m=-1$ mode, especially at higher $Ha$. 

Figure \ref{fig:amriconvectiongrowthvelo}(b) shows the normalised energy content $\rm A^2$ of perturbations as obtained both from experiments and the linear stability analysis for the $m=\pm 1$ AMRI modes when the heat flux is directed from the inner to outer cylinder. The experimental data for $\rm A^2$ represents the square of the measured rms of the axial velocity induced by AMRI, which are normalised by their maximum value with respect to $Ha$. On the other hand, the theoretical values are assumed to be proportional to the growth rate $\gamma_a$ of absolute AMRI, i.e., $A^2\propto \gamma_a$, as is typical of a slightly supercritical regime according to Ginzburg-Landau theory of weakly nonlinear processes \citep[e.g.,][]{Landau,Umurhan_etal2007}.  The theoretical values of $\rm A^2$ are also normalised by their maximum over $Ha$ corresponding to the $m=1$ AMRI mode, which is the dominant mode in this case, since its growth rate is enhanced by convection, as is seen in figure  \ref{fig:amriconvectiongrowthvelo}(b). This enhancement of the growth rate is due to the additional free energy provided selectively for the $m=1$ mode by the shear of the axial convective velocity, $u_{0z}'$. Such a normalization allows us to compare the relative magnitudes (growth rates) of the $m = \pm 1$ AMRI modes and the form of their dependence on $Ha$ in the presence and absence of convection as well as with the experimental data. Namely, the $m=1$ AMRI modes are the strongest, while the $m=-1$ AMRI modes the weakest, with the pure AMRI modes being between these two. This clearly demonstrates the nature of symmetry breaking between the $m=\pm 1$ modes caused by $u_{0z}$ -- increase in the amplitude (growth rate) of one ($m=1$) mode and decrease that of another ($m=-1$). Note also that the critical $Ha_c\approx 62$ is lower than $Ha_c\approx 82$ for the AMRI without convection. 

Thus, it is evident from figure \ref{fig:amriconvectiongrowthvelo} that the experimental data are in a good agreement with the theoretical results for the dominant $m=1$ AMRI mode for the radially outward heat flux both for the phase velocity and energy content, especially near the onset at $62\lesssim Ha\lesssim 100$, where the data points are closest to the $m=1$ mode curve.  
%%%===============================================%%%
%%%===============================================%%%
%\vspace{-1em}
\section{Conclusion}
%%%===============================================%%%
%%%===============================================%%%

In this paper, we performed linear stability analysis for the absolute version of AMRI in the presence of thermal convection and compared it with the experimental results from PROMISE. 
The theoretical prediction of the early onset of AMRI and symmetry breaking between $m=\pm 1$ modes brought about by convection are in a good agreement with the experimental results. This symmetry breaking is manifested in the increase in the growth rate of either of these two modes with a given $m$ and decreases that of another. As a result, AMRI sets in at lower critical $Ha_c$ than that in the absence of convection. Although the experimental and theoretical results are consistent in yielding the dependence of the amplitudes of the AMRI waves on Hartmann number based on the compariosn with the measured axial convective velocity, future simulations are needed to obtain these features taking into account nonlinear feedback of AMRI on the convection. 

Our findings may have implications for a subcritical MRI-dynamo in astrophysics \citep{Rincon2019}, which is sensitive to the amplitude of initial perturbations and generally involves non-axisymmetric modes. Specifically, in the solar tachocline where $Pm\ll 1$, the combination of differential rotation and thermal convection may foster AMRI modes with large enough amplitudes to sustain AMRI-driven dynamo. Furthermore, our preliminary WKB analysis indicates that $|m|\geq 2$ modes can be unstable at higher $Ha$ and/or $Re$ that may be of astrophysical importance. New experiments with an upgraded PROMISE setup including thermal processes may be helpful in this respect.
\\
\\
We thank Prof.  R.  Hollerbach for providing the
linear 1D code. This work received funding from the European Union’s Horizon 2020 research and innovation program under the ERC Advanced Grant Agreement No. 787544 and from
Shota Rustaveli National Science Foundation of Georgia
(SRNSFG) [grant number FR-23-1277].
\\
\\
\textbf{Declaration of Interests:} The authors report no conflict of interest.

\bibliographystyle{jfm}
% % Note the spaces between the initials
\bibliography{jfm-instructions}

\end{document}